\documentclass[preprint,12pt]{elsarticle}

\usepackage{amssymb}
\usepackage{graphics}
\usepackage{graphicx}
\usepackage{epsfig}
\usepackage{amssymb}
\usepackage{amsfonts}
\usepackage{color}

\journal{Physics Letters A}

\begin{document}

\begin{frontmatter}



\title{Renormalization group approach to a  $p$-wave superconducting model}


\author{Mucio A. Continentino}
\author{Fernanda Deus}

\address{Centro Brasileiro de Pesquisas Fisicas, Rua Dr. Xavier Sigaud, 150, Urca \\
22290-180, Rio de Janeiro, RJ, Brazil}
\author{Heron Caldas}
\address{Departamento de Ci\^{e}ncias Naturais, Universidade Federal de
  S\~ao Jo\~ao Del Rei, \\ 36301-000, S\~ao Jo\~ao Del Rei, MG, Brazil}

\begin{abstract}
We present in this work an exact renormalization group (RG) treatment of a one-dimensional $p$-wave superconductor. The model proposed by Kitaev consists of a chain of spinless fermions with a $p$-wave gap. It is a paradigmatic model of great actual interest since it presents a weak pairing superconducting phase that has Majorana fermions at the ends of the chain. Those are predicted to be useful for quantum computation. The RG allows to obtain the phase diagram of the model and to study the quantum phase transition from the weak to the strong pairing phase. It yields the attractors of these phases and the critical exponents of the weak to strong pairing transition. We show that the weak pairing phase of the model is governed by a chaotic attractor being non-trivial from both its topological and RG properties. In the strong pairing phase the RG flow is towards a conventional strong coupling fixed point.  Finally, we propose an alternative way for obtaining  $p$-wave superconductivity in a one-dimensional system  without  spin-orbit interaction.

\end{abstract}


\end{frontmatter}


\section{Introduction}

Majorana fermions~\cite{majorana} at the end of $p$-wave superconducting wires are promising quasi-particles to act as qubits in
quantum computers~\cite{quantum}. Those particles are topologically protected satisfying a criterion of robustness required for quantum computation. On the other hand they have an interest in themselves as new particles with exotic properties~\cite{majorana}. A model proposed by Kitaev~\cite{kitaev} was shown to exhibit these Majorana fermions. In spite of its apparent simplicity, it 
is a paradigmatic model for $p$-wave superconductors exhibiting all the complexity of topological phases, edge states and Majorana fermions~\cite{read}. Besides, it offers the possibility to be realized in practice in actual physical systems~\cite{sarma,alicea}. A full understanding of its properties  including the nature of its phases and its critical behavior is essential to make progresses on these important topics.
In this letter we investigate the Kitaev model using a renormalization group (RG) approach and discuss a new possibility for realizing it experimentally.

The Kitaev model consists of  a chain with spinless fermions and an attractive interaction that gives rise to a $p$-wave superconducting gap~\cite{kitaev}. This $k$-dependent gap vanishes at points of symmetry that correspond to the bottom and top of a fermionic band. At zero temperature this model presents two superconducting phases.  For weak couplings there is a  {\it weak pairing} phase that is topologically non-trivial and contains Majorana fermions at the ends of the chain. This phase has a double degenerescence that can be associated with the presence of the Majorana particles. The other phase is  a {\it strong pairing} phase which is  topologically trivial  and has a unique ground state. Although these phases have been characterized from the point of view of their topological properties~\cite{read,alicea} to the best of our knowledge there is no RG study of this model. Here we show that the weak pairing phase besides its nontrivial topological character is also nontrivial from the RG point of view since it is associated with a {\it chaotic attractor}~\cite{chaos}. 
On the other hand, in the strong pairing phase the RG equations have a conventional behavior and all points in this phase iterate to a strong coupling attractor.
As a consequence of our analysis we find a correspondence between a non-trivial topological phase and a non-trivial renormalization group description of this phase. The opposite is also true for the trivial strong pairing phase that is governed by a conventional strong coupling attractor.

The RG approach allows to fully characterize the universality class of the quantum phase transition between the two superconducting phases. This weak-to-strong pairing transition is associated with a fully unstable fixed point at the top of the conduction band. The flows of the RG equations close to this fixed point allow to obtain the correlation length and the dynamic exponents characterizing this transition. 

Finally for completeness we present a new multi-band one-dimensional (1D) model that exhibits $p$-wave superconductivity. The model differently from those which appear in the literature of this problem based on the spin-orbit interaction, rely on an odd-parity hybridization between orbitals of different parities on neighboring sites of the chain.

\section{Hamiltonian}
We consider a linear chain with spinless fermions that can hop to nearest neighbor sites and have an attractive interaction that gives rise to an
odd  pairing gap $\Delta_{ij}$.
The Kitaev model~\cite{kitaev} in real space can be written as,
\begin{equation}
\mathcal{H}=-\frac{1}{2} \sum_{ij} t_{ij} c^{\dagger}_i c_{j}  -\frac{1}{2}  \sum_{ij} \left( \Delta_{ij} c^{\dagger}_i c^{\dagger}_j +  \Delta^{*}_{ij} c_i c_j \right) -\mu \sum_{i} n_i
\end{equation}
where $t_{ij}$ is a nearest neighbor hopping, $\mu$ the chemical potential and $\Delta_{ij}=-\Delta_{ji}$ an odd pairing between fermions in neighboring sites. The operators $c_i$ and $c_i^{\dagger}$ destroy and create fermions on site $i$ of the chain, respectively. 
Fourier transforming this Hamiltonian we obtain,
\begin{equation}
\mathcal{H}=\sum_{k} (\epsilon_k -\mu) c^{\dagger}_k c_{k} - \sum_{k} \left( \Delta_{k} c^{\dagger}_k c^{\dagger}_{-k} +  \Delta^{*}_{k} c_{-k} c_k \right)
\end{equation}
where,
\begin{eqnarray}
\label{basic1}
\epsilon_k &=& -t \cos k a \\
\Delta_{k}&=& - i \Delta_0 \sin k a. \label{basic2}
\end{eqnarray}
Notice that $a$ is the distance between sites on the chain and $\Delta_0$ is a complex constant. 

We now perform a renormalization group transformation  removing every other site in the chain. The new lattice spacing is $a^{\prime}=a/2$. In momentum space this corresponds to take $k^{\prime} = 2 k$. Here we apply  the renormalization group transformation in momentum space~\cite{oliveira,livro}. In the renormalized lattice $k^{\prime}$ replaces $k$ in Eqs.~\ref{basic1} and~\ref{basic2}. Using $k^{\prime} = 2 k$ and the relations,
\begin{eqnarray}
\label{cos}
\cos k^{\prime} a = \cos 2 k a = 2 \cos^2k a -1 \\
\label{sin}
\sin k^{\prime} a = \sin 2 k a = 2 \sin k a \cos k a,
\end{eqnarray}
we obtain a new Hamiltonian with the same form as the previous one, but with renormalized parameters given by,
\begin{eqnarray*}
\Omega^{\prime} &=& \Omega^2 -2 \\
\delta^{\prime} &=& 2 \delta \sqrt{1+\delta^2} 
\end{eqnarray*}
where  $\Omega=-(\omega_k+ \mu)/(t/2)$, $\delta = \Delta / \Delta_0$ and we defined $\omega_k= \epsilon_k - \mu$. 
We can write these 
equations in the form of recursion relations as, 
\begin{eqnarray}
\label{RG1}
\Omega_{n+1} &=& \Omega_n^2 -2   \\
\delta_{n+1} &=& 2 \delta_n \sqrt{1+\delta_n^2} \label{RG2}
\end{eqnarray}

Let us consider  Eqs.~\ref{RG1} and~\ref{RG2} and their fixed points. Eq.~\ref{RG1} is known as the logistic map~\cite{ulam,feigenbaum}. It has two unstable fixed points at $\Omega^{*}=2$ and $\Omega^{*}=-1$. The former divides the $\Omega$ axis in two distinct regions: the region $|\Omega| > 2$ where all initial points $|\Omega_0|>2$ iterate to infinity under successive renormalization group transformations,  and the region $|\Omega| < 2$  where any initial  point $|\Omega_0| < 2$ remains always in this interval under iteration. The unstable fixed point at $\Omega^{*}=2$ corresponds to the bottom of the band at $k=0$. The point $\Omega_0 = -2$ iterates to the fixed point $\Omega^{*}=2$ and corresponds to the top of the fermion band. The renormalization group equation~\ref{RG1} in the region $[-2,2]$ has periodic orbits but is chaotic for most of the initial points, as any point in this interval  is reached arbitrarily close if the system is iterated a sufficiently large number of times~\cite{feigenbaum}.
The  fixed point, $\Omega^{*}=-1$ and the point $\Omega=1$ that maps into $\Omega^{*}=-1$  correspond to values of $k$, such that, $\cos ka =\pm 1/2$, i.e. to $ka=\pm \pi/3$ and $ka=\pm 2\pi/3$.

Next we analyze Eq.~\ref{RG2}. The fixed points occur for $\delta^{*}=0$ and $\delta^{*}=\pm i \sqrt{3/4}$. The former is unstable and corresponds to $ka=0$ and $ka=\pi$, the bottom and the top of the band of fermions. The latter are also unstable and correspond to $ka=\pm \pi/3$  and $ka=\pm 2\pi/3$. The values of $\delta$ iterate always in the interval ($-i,i$), such that, the  gaps generated by the RG procedure in this region are always smaller than the initial gap $\Delta_0$. 

\begin{figure}[th]
\centering{\includegraphics[scale=0.8]{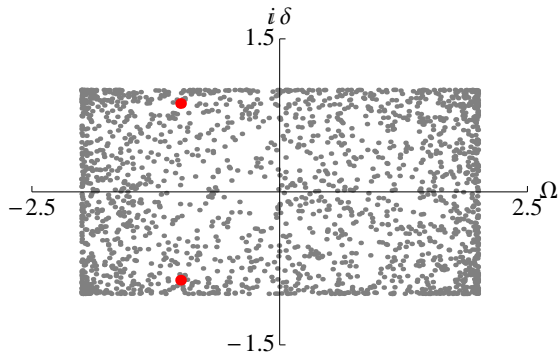}}\caption{(Color online) The weak pairing phase which is generated by iterating the RG equations (Eqs.~\ref{RG1} and~\ref{RG2}) in the neighborhood of the fixed points  ($\Omega^{*}, \delta^{*}$)=($-1, \pm  i \sqrt{3/4}$) shown as big dots.}%
\label{fig1}%
\end{figure}

Summarizing, Eqs.~\ref{RG1} and~\ref{RG2} have unstable fixed points at the bottom of the conduction band $k=0$ corresponding to ($\Omega^{*}, \delta^{*}$)=($2,0$). The states at the top of the band, $k=\pi /a$ map to these fixed points. The fixed points ($\Omega^{*}, \delta^{*}$)=(-1,$\pm i \sqrt{3/4}$) correspond to values of $ka=\pm \pi/3$  and $ka=\pm 2\pi/3$ inside the band of conduction states. The iteration of Eqs.~\ref{RG1} and ~\ref{RG2} with initial points in the neighborhood of these fixed points gives rise to a chaotic sequence where all points in the interval $\Omega \ni [-2,2]$, $\delta \ni [-i,i]$ are visited arbitrarily close for a sufficient large number of iterations (see Fig~\ref{fig1}) for almost all initial values. 


The RG equations in the form above are independent of each other. However, to describe appropriately the system we must consider a coupling of the equations for the kinetic energy and the gap since they are naturally constrained as will become clear below. This coupling arises, for example, from the relations Eqs.~\ref{cos} and\ref{sin}. It constrains at least one of the recursion relations that can now be written as, 
\begin{eqnarray}
\label{RGC1}
\Omega_{n+1} &=& \Omega_n^2 -2  \\
\label{RGC2}
\delta_{n+1} &=& \Omega_n \delta_n 
\end{eqnarray}
These equations have a structure of fixed points similar to that of Eqs.~\ref{RG1} and~\ref{RG2} that have been analyzed before. There is an unstable fixed point at $\Omega^{*}=2, \delta^{*}=0$, that corresponds to $k=0$ at the bottom of the fermion band. The fixed points $\Omega^{*}=-1, \delta^{*}=\pm i \sqrt{3/4}$  correspond to $ka=\pm 2\pi/3$. They iterate, such that, if we start with one of them, say  $\Omega^{*}=-1, \delta^{*}=i \sqrt{3/4}$, we obtain the sequence ($-1,- i \sqrt{3/4}$),  ($-1, i \sqrt{3/4}$),  ($-1,- i \sqrt{3/4}$),   $\cdots$, showing that these fixed points represent a periodic orbit of period $2$. They are associated with the degeneracy of the ground state as they yield the same values for $|\delta|$. 
This double degeneracy of the weak pairing phase arises from the  presence of two Majorana fermions in the ends of the chain~\cite{alicea}.

The next periodic orbit is period $4$ that iterates to pairs $(\Omega^{*}, \delta^{*})$ given by,  ($-1.618,   0.587$), ($0.618, -0.951$), 
($-1.618,  -0.587$), ($0.618,   0.951$), ($-1.618$, $0.587$). These states however are not degenerate as they give rise to different values of the gap.

The coupled RG equations, Eqs.~\ref{RG1} and~\ref{RG2} allow now to obtain the phases and the critical behavior of the Kitaev model.
When the chemical potential is such that $|\Omega|> 2$, $\delta > 0$, the recursion relations, Eqs.~\ref{RGC1} and~\ref{RGC2} iterate to the strong coupling fixed point ($\infty$, $i$$\infty$). This attractor characterizes the {\it strong pairing} phase that is a trivial superconductor with no special topological properties. Then, the  same conventional behavior appears both in the topological properties and in RG description of this phase.

\begin{figure}[th]
\centering{\includegraphics[scale=0.8]{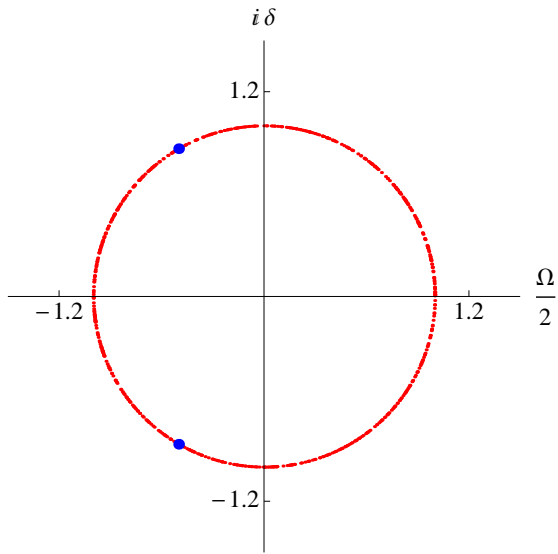}}\caption{(Color online) Iteration of Eqs.~\ref{RGC1} and~\ref{RGC2} starting from $\Omega=\Omega_0$ ($|\Omega_0| <2$) and $i\delta=\sqrt{1-\Omega_0^2/4}$. The equation for the attractor
 where the points iterate is given by, $\Omega^2/4 +(i \delta_n)^2=1$. The period $2$ orbit at $\Omega^{*}=-1, \delta^{*}=\pm i \sqrt{3/4}$ is also shown as large dots.}%
\label{fig2}%
\end{figure}

When the chemical potential is such that $\Omega$ lies in the interval $[-2,2]$ and $\delta > 0$, the behavior of the RG equations is chaotic~\cite{logistic} since it is governed by Eq.~\ref{RGC1}. Using that at each step of the renormalization procedure, $\Omega_n/2=\cos k_n a$ and $ (i \Delta_n/\Delta_0)= \sin k_n a$ we square and add these equations to obtain,
\begin{equation}
\label{attractorweak}
\Omega_n^2/4 +(i \delta_n)^2=1.
\end{equation}
Then most of the initial points belonging to this line (this excludes those points that give rise to periodic orbits or that iterated directly to the fixed points)  iterate chaotically but always remain constrained to it.  This is shown in Fig.~\ref{fig2} and also the period $2$ orbit at $\Omega^{*}=-1, \delta^{*}=\pm i \sqrt{3/4}$. In particular initial points in the neighborhood of this orbit visit arbitrarily close any point of the curve for a sufficiently large number of iterations. Notice that $\delta$ has the maximum value of unity implying that the gap always iterate to values smaller than the amplitude $\Delta_0$, whenever $\Omega$ lies in the interval $|\Omega| <2$. This {\it weak pairing} phase differently from the usual RG description, as that of the strong pairing phase, is not characterized by an attractive fixed point  but by a chaotic attractor which is shown in Fig.~\ref{fig2}, and given  by Eq.~\ref{attractorweak}. Furthermore this curve, Eq.~\ref{attractorweak}, contains a periodic orbit of period $2$ that we associate with the double degenerescense of this phase.
Thus the non-trivial topological properties of the weak pairing phase of the Kitaev model has a counterpart also in its renormalization group description.

Due to the constraint Eq.~\ref{attractorweak}, the attractor of the weak pairing phase can be described in terms of a single recursion relation, namely
\begin{equation}
\label{circlemap}
\theta_{n+1}= 2 \theta_n,
\end{equation}
modulo 1, with the arguments of the trigonometric function being replace by, $ka \rightarrow 2 \pi \theta$.
Eq.~\ref{circlemap} is that for the circle map, which has been intensively studied in the theory of chaotic systems~\cite{chaos}. 

\begin{figure}[th]
\centering{\includegraphics[scale=0.8]{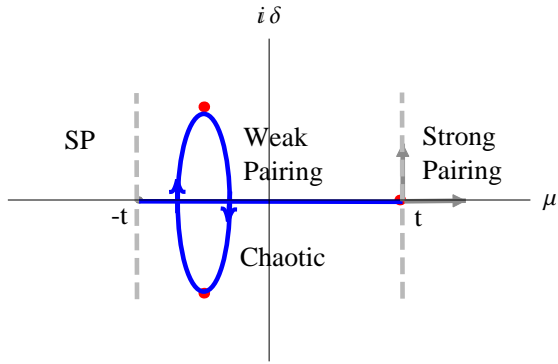}}\caption{(Color online) The phase diagram of the Kitaev model, showing the weak pairing (chaotic) and strong pairing phases, the unstable fixed points and the flow of the RG equations (arrows).}%
\label{fig3}%
\end{figure}

The phase diagram of the model is shown in Fig.~\ref{fig3}. As the chemical potential decreases the system goes from the strong pairing to the weak pairing phase.
The phase transition from the strong  to the weak  pairing phase is governed by the unstable fixed point  $\Omega^{*}=2, \delta^{*}=0$. In the non-interacting case, the fixed point at $\Omega^{*}=2, \delta^{*}=0$ governs the density-driven or Lifshitz transition metal-insulator transition. This is the simplest case of a phase transition in a non-interacting fermionic system.

When turning on the pairing interaction, we find it is a relevant perturbation at the fixed point $\Omega^{*}=2, \delta^{*}=0$.  The quantum critical exponents associated with this quantum critical point (QCP) in the presence of interactions can be obtained from the Jacobian of the RG transformations, we get.

\[
J=\left(
\begin{array}{cc}
  \frac{\partial \Omega_{n+1}}{\partial \Omega_n}   &   \frac{\partial \Omega_{n+1}}{\partial \delta_n} \\ 
    \frac{\partial \delta_{n+1}}{\partial \Omega_n}   &   \frac{\partial \delta_{n+1}}{\partial \delta_n} \\ 
  \end{array}
\right)
\]
At the QCP ($\Omega^{*}=2, \delta^{*}=0$), this yields
\[J=
\left(
\begin{array}{ccc}
  4   &  0 \\
 0  &  2 \\
\end{array}
\right)
\]
Since this is diagonal it implies that the two relevant directions ($\Omega$ and $\delta$)  are orthogonal. The critical exponents are obtained from the eigenvalues $\lambda_1=4$ and $\lambda_2=2$. The gap at the quantum critical point $\Omega^{*}=2, \delta^{*}=0$ scales as $\delta^{\prime}= b^z \delta$ which defines the dynamic quantum critical exponent. Using that $b^z=\lambda_2=2$ and the scaling factor $b=2$ we obtain for the dynamic exponent the value $z=1$. On the other hand, for $\delta=0$, we expand the RG equations close to $\Omega^{*}$ and obtain, $\Omega^{\prime}= \Omega^{*} + b^{1/\nu} (\Omega-\Omega^{*})$. Since $b^{1/\nu}=\lambda_1=4$, using $b=2$  we identify the correlation length exponent $\nu=1/2$. The crossover exponent $\phi=\lambda_2/\lambda_1 = \nu z =1/2$. Notice that the  shift exponent~\cite{livro} that determines the semi-circular shape of the boundary of the weak pairing phase (see Fig.~\ref{fig2}) coincides with the crossover exponent, i.e., $\psi=\nu z=1/2$. These values for the exponents determine the universality class of the quantum phase transition from the weak pairing to the strong pairing superconductor. They are consistent with the form of the excitation spectrum as discussed below.
Notice that in the present problem instead of having a smooth crossover from a weak to a strong coupled superconductor as usual for $s$-wave superconductors~\cite{becbcs} we have a true quantum phase transition separating a weak from a strong pairing phase. Since both phases are superfluid, the nature of the phase transition is topological. However, from the RG perspective this unconventional transition is still associated with an unstable fixed point and has well defined critical exponents.

\subsection{Excitation spectrum}

The energy of the excitations in the superconducting phase is given by,
\begin{equation}
\omega(k)= \sqrt{ (-\mu - t \cos ka)^2 + |\Delta_0|^2 \sin^2 ka}
\end{equation}
This equation has zero modes for $\mu= \pm t$ for $ka=0$ and $ka=\pi$, i.e, when the chemical potential is at the border of the conduction band. This zero energy mode is required since the systems passes from a topological weak pairing phase to a trivial, non-topological strong pairing phase. Close to the phase transition the spectrum can be written as
\begin{equation}
\omega(k)= \sqrt{(\mu-\mu_c)^2 +|\Delta_0|^2 a^2 k^2}
\end{equation}
where $|\mu_c|=t$. Then the spectrum is Dirac-like at the QCP, $\mu=\mu_c$ with a velocity $|\Delta_0| a$. This linear spectrum is related to the value of the dynamic exponent $z=1$ found before using the RG procedure.

\begin{figure}[th]
\centering{\includegraphics[scale=0.8]{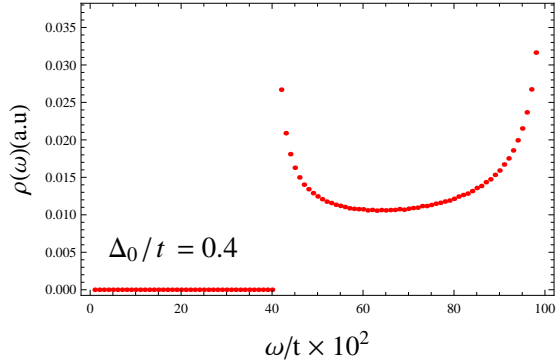}}\caption{(Color online) The density of states $\rho(\omega)$ in arbitrary units (a.u)  for the excitations of the Kitaev model in the weak pairing phase for $\Delta_0/t=0.4$.}
\label{fig4}
\end{figure}

The density of states $\rho(\omega)$ of the {\it bogoliubons}, i.e., the excitations in the weak pairing phase can be obtained from the density of visits of the recursion relations in a given energy interval~\cite{oliveira}.  Explicitly, we write ($\mu=0$),
\begin{equation}
\label{ro}
f(\Omega_n)=\sqrt{\frac{\Omega_n^2}{4}\left(1-(\frac{\Delta_0}{t})^2\right)+(\frac{\Delta_0}{t})^2}
\end{equation}
where $\Omega_n$ iterates according to Eq.~\ref{RG1}. 
We iterate this equation a large number of times ($6 \times 10^6$),  substitute the generated values of $\Omega_n$ in Eq.~\ref{ro} and count the {\it number of visits} of $f(\Omega_n)$ in a given energy interval $\omega + d\omega$. The density of states $\rho(\omega)$ of the excitations is proportional to the number of visits of $f(\Omega_n)$ in this interval~\cite{oliveira}. The result of this procedure is shown in Fig.~\ref{fig4}. 

\section{Two-band model with anti-symmetric hybridization} 
 
 \quad In this section we discuss a possible realization of a 1D system with the main characteristics of the Kitaev model discussed above. Most of the proposals to obtain in practice a $p$-wave onde dimensional superconductor have relied on the spin orbit interaction as a main ingredient to confer odd parity to the order parameter~\cite{alicea,soc}. Here we propose an alternative that consists of a   two-band system  with hybridization between these bands and an attractive interaction between them~\cite{recent}.
The hybridization occurs due to the mixing of different orbitals in neighboring sites by the crystalline potential and consequently it is $k$-dependent. Most important we consider orbitals with different parities, such as, orbitals with angular angular $l$ and $l+1$, as $sp$, $pd$ or $df$ orbitals.  This guarantees that the $k$-dependent hybridization has odd parity ($V(-k)=-V(k)$). For the 1D case considered here $V(k)=2i V_0 \sin ka$ where $a$ is the lattice spacing. The attractive inter-band interaction is treated in the BCS approximation, and now differently from the previous sections it is considered to be $k$-independent. For concreteness we consider a block of a $d$-metal superconductor as $Nb$ on top of which is deposited a wire of a $p$-metal superconductor as $In$ or $Sn$. The relevant part of the Hamiltonian is given by,

\begin{eqnarray}
\mathcal{H}=\sum_{k\sigma} \left( \epsilon^{a}_{k}a^{\dag}_{k\sigma}a_{k\sigma}+\epsilon^{b}_{k}b^{\dag}_{k\sigma}b_{k \sigma} \right)- \sum_{k\sigma} \left( \Delta_{bb} b^{\dag}_{k\sigma}b^{\dag}_{-k -\sigma}+\Delta_{bb}^* b_{k-\sigma}b_{k\sigma}\right)\\
-\sum_{k\sigma} \left( \Delta_{ab} a^{\dag}_{k\sigma}b^{\dag}_{-k -\sigma}+\Delta_{ab}^* b_{k-\sigma}a_{k\sigma}\right)
+\sum_{k\sigma}\left(V_{k}  a^{\dag}_{k \sigma}b_{k \sigma}+ V^{*}_{k}b^{\dag}_{k\sigma}a_{k \sigma} \right)
 \label{eq1}
\end{eqnarray}%
where $\epsilon^{a,b}_{k}=-t_{a,b} \cos ka-\mu_{a,,b}$ are the energies of the electrons in the $a$ and $b$ bands (the $p$ and $d$ bands of the $Sn$ wire and  of the bulk $Nb$, respectively). In an obvious notation $a^{\dagger}_{k \sigma}$ and  $b^{\dagger}_{k \sigma}$ create electrons in these bands respectively.  As pointed out before, we consider that the orbitals $a$ and $b$ have different parities as for states with orbital angular momentum $l$ and $l+1$ ($p$ and $d$ in the present case). In this case the $k$-hybridization for the linear chain $V_k = 2 i V_0 \sin ka$ has odd parity. This is due to the fact that the mixing between these different orbitals in neighboring sites is such that $V_{ij}=-V_{ji}$.  Notice that hybridization, in contrast to the spin orbit interaction mixes electrons with the {\it same} spin and does not require breaking of inversion symmetry of the lattice itself. 

The order parameters that characterizes the superconducting phase are given by,
\begin{eqnarray}
\label{eq2}
\Delta_{bb}=g_{bb}\sum_{k\sigma}\langle b_{-k-\sigma}b_{k\sigma}\rangle\\
\Delta_{ab}=g_{ab}\sum_{k\sigma}\langle b_{-k-\sigma}a_{k\sigma}\rangle
\label{eq3}
\end{eqnarray}
where $g_{bb}$, $g_{ab}>0$ are coupling constants.

The energy of the excitations in the superconducting phase are given  by,  $\pm \omega_{1,2}(k)$, where,
\begin{eqnarray}
\omega_{1}=\sqrt{A_k + \sqrt{A_k^2- R_k}} = - \omega_{3} \\
\omega_{2}=\sqrt{A_k - \sqrt{A_k^2-R_k}} = - \omega_{4}
\end{eqnarray}
with
\begin{equation}
\label{akt}
A_k=\frac{ \epsilon_k^{a2}+ \epsilon_k^{b2}}{2} +\frac{|\Delta_{bb}|^2}{2}+ |\Delta_{ab}|^2+|V_k|^2
\end{equation}
and
\begin{eqnarray}
R_{k} =  \left( |V_k|^2 -(\Delta_{ab}^2 +\epsilon_{k}^{a} \epsilon_{k}^{b}) \right)^2 +(\Delta_{bb} \epsilon_{k}^{a}-2 V_k \Delta_{ab})^2
\label{bkt}.
\end{eqnarray}

The self-consistent equations for the order parameters can be easily obtained from the anomalous Green's functions, 
\begin{equation}
\label{anomalous}
\mathcal{F}_{\alpha \beta}(k,\omega)=\frac{1}{2 \pi}\frac{f_{\alpha \beta}(k,\omega)}{(\omega^2-\omega_1^2)(\omega^2-\omega_2^2)},
\end{equation}
with $\alpha \beta = bb$ and $ab$, using the fluctuation-dissipation theorem. It turns out that in spite of the absence of an attractive intra-band interaction in the $a$-band, there are anomalous superconducting correlations $ \langle a_{k \sigma}a_{-k -\sigma}\rangle$ in this band  induced by hybridization and/or inter-band interactions. They can be obtained from an anomalous Green's function $\mathcal{F}_{aa}(k, \omega)$ as in Eq.~\ref{anomalous}, where $f_{aa}$ is given by the frequency independent function,
\begin{equation}
\label{aux}
f_{aa}(k)=-|V_k|^2 \Delta_{bb}+\Delta_{bb} \Delta_{ab}^2+2 V_k \epsilon_k^b \Delta_{ab}.
\end{equation}
Then using Eqs.~\ref{anomalous} and~\ref{aux} and the fluctuation-dissipation theorem we obtain the anomalous correlation function $\langle a_{k \sigma} a_{-k -\sigma}\rangle$. This has  three contributions. The first proportional to $|V_k|^2$ is the usual induced superconductivity due to the {\it proximity effect} that arises in a metal in close contact with a superconductor. The next two contributions require an attractive interaction $g_{ab}$ between the electrons in the wire and those in the bulk superconductor, such that, $\Delta_{ab} \ne 0$. Notice that the last term, because of  the odd parity of $V_k \propto \sin ka$, induces a $p$-wave type of superconductivity in the wire ( the $a$-band system). This term can become dominant if $|V_k|^2 \approx \Delta_{ab}^2$.

The induced $p$-wave type contribution for the  anomalous correlations function $\langle a_{k \sigma}a_{-k -\sigma}\rangle$ vanishes at $k=0$, as expected from a $p$-wave superconductor, and at the Fermi surface of the original $b$-band, since  $ \epsilon^b_{k_F^a}=0$. 

\section{Conclusions}

We have studied Kitaev $p$-wave 1D superconducting model~\cite{kitaev} using an exact renormalization group approach. We obtained the zero temperature phase diagram of the model and the universality class of the weak-to-strong pairing quantum phase transition. We have shown that the non-trivial topologically weak pairing phase is governed by a chaotic attractor with a bistable fixed point. The strong pairing phase with trivial topological properties is governed by a conventional strong coupling fixed point.
We have proposed an alternative  way for obtaining $p$-wave superconductivity in a wire by considering instead of the spin-orbit interaction, an odd-parity hybridization between  the orbitals of the wire and those of the bulk superconducting metal on top of which the wire is deposited. The angular momentum of the orbitals of the wire must differ by one, such that, they have different parities and hybridize anti-symmetrically.  As shown by Kitaev~\cite{kitaev} $p$-wave superconducting  chains  as those studied here exhibit Majorana quasi-particles at their ends. These are expected to be useful for quantum computation.

Acknowledgements: MAC would like to thank A. M. Ozorio de Almeida for useful discussions.   We wish to thank the Brazilian agencies, CAPES, 
FAPERJ, CNPq and FAPEMIG for financial support. HC acknowledges the hospitality of CBPF where this work was done.

\end{document}